\documentclass{article}
\usepackage{amssymb}
\usepackage{amsmath}
\usepackage{graphicx}
\usepackage{amsthm}

\usepackage{relsize}

\def \mc {\mathcal}
\def \d {\mathrm{d}}

\def \ni {\noindent}

\begin{document}
\title{\bf{\Large Averaging geometrical objects on a differentiable manifold}}
\author {{\small Johan Brannlund \footnote{johanb@mathstat.dal.ca}} \\
\it{\small Department of Mathematics and Statistics, Dalhousie University} \\
\it{\small Halifax, Nova Scotia, B3H 3J5, Canada }
\and
{\small Robert van den Hoogen \footnote{rvandenh@stfx.ca}} \\
\it{\small Department of Mathematics, Statistics and Computer Science} \\
\it{\small St Francis Xavier University} \\
\it{\small Antigonish, Nova Scotia, B2G 2W5, Canada}
\and
{\small Alan Coley \footnote{aac@mathstat.dal.ca}} \\
\it{\small Department of Mathematics and Statistics, Dalhousie University} \\
\it{\small Halifax, Nova Scotia, B3H 3J5, Canada }
}
\maketitle
\begin{abstract}

  We construct a framework within which a mathematically precise,
  fully covariant, and exact averaging procedure for tensor fields on
  a manifold can be formulated. In particular, we introduce the
  Weitzenb\"ock connection for parallel transport and argue that,
  within the context of averaging, frames and connections are the
  natural geometrical objects on the manifold.
\end{abstract}

\newpage

\section{Introduction}

In order to address several physical problems in gravitational
theories, including the important question of averaging in cosmology
and the interpretation of cosmological observations \cite{Coley}, it
is necessary to define an averaged or macroscopic theory of
gravity. However, due to the non-linear nature of the gravitational
field equations in, for example, Einstein's General Relativity (GR),
it is very difficult to define a mathematically precise and covariant
averaging procedure for tensor fields. In previous approaches a $3+1$
cosmological space-time splitting has been employed (i.e., this
procedure is not generally covariant) where only scalar quantities are
averaged \cite{Buch} and a perturbative approach involving averaging
the perturbed Einstein Field Equations (EFE) has been utilized
\cite{Kolb}.

In the macroscopic gravity (MG) approach to the averaging problem in
GR, which is a fully covariant and exact method, a prescription for
the correlation functions which emerge in an averaging of the
non-linear field equations (without which, the averaging of the EFE
simply amount to definitions of the new averaged terms) is given, and
a new tensor field with its own set of field equations is introduced
\cite{Zal}.  For the cosmological problem, additional assumptions are
required: with reasonable cosmological assumptions, the correlation
tensor in MG takes the form of a spatial curvature \cite{CPZ}. The
formal mathematical issues of averaging tensors on a differential
manifold have recently been revisited \cite{other,vdh2010}. In
particular, we note that it may be possible to avoid several of the
technical problems of averaging by adopting an approach based on
scalar curvature invariants \cite{Coley}.

The aim of this paper is to construct a mathematical framework within
which a consistent theory of averaged geometric quantities on a
differentiable manifold can be formulated. While we develop the
formalism with applications to GR in mind (applications to cosmology
will be pursued in future work {\footnote{ On small enough scales, our
    universe is inhomogeneous and it is on these small scales that the
    theory of GR has been tested. One may therefore adopt the point of
    view that we should start with GR in the small and average out the
    inhomogeneities to obtain an effective large-scale theory. The
    non-linear nature of the EFE means that this large-scale theory is
    not necessarily GR itself.}}), our primary motive is to set up a
smoothing procedure of interest in its own right, similar in spirit to
concepts such as Ricci flow \cite{ham}.

Our intent is to introduce a rigorously defined geometric averaging
procedure that captures the intuition of averaging as integrating a
quantity over a region $\Sigma$ and dividing by the volume
$V_\Sigma$. When we average a tensorial object $T$, in order for the
averaged object $\overline{T}$ to transform tensorially, $T$ has to be
parallel transported from each point $x'$ within $\Sigma$ to a common
reference point\footnote{We will sometimes write $\Sigma_x$ when
we want to indicate the reference point.} $x$. 
Therefore, as in \cite{vdh2010}, we introduce the
Weitzenb\"ock connection for parallel transport and argue that, within
the context of averaging, frames and connections are more natural
geometrical objects on the manifold. This also suggests alternative
theories of gravity (such as, for example, Poincar\'e Gauge Theories)
as being of interest.

\begin{figure}[h]
  \centering
\includegraphics[height=55mm]{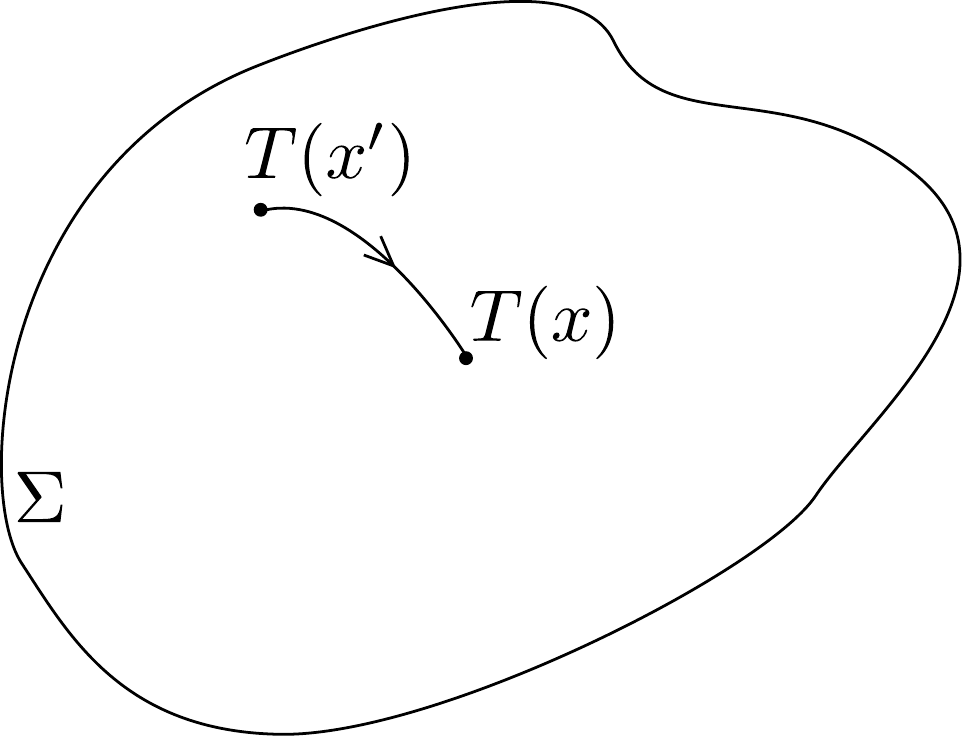}
\caption{$T(x')$ being parallel transported to a reference point $x$, in order
to define the average over $\Sigma$.}
\end{figure}

\section{Mathematical Framework}

We will assume a differentiable $p$-dimensional manifold $\mc{M}$,
equipped with a local frame co-basis $e_a^I$ whose associated metric
$g_{ab}=e_a^Ie_{bI}$ is of Lorentzian or Riemannian signature and a
connection form $\omega^I_{Ja}$ taking values in the Lie algebra
$\mathfrak{g}$ of the appropriate pseudo-orthogonal structure group
$G$, which is $\textrm{SO}(p-1,1)$ for Lorentzian manifolds and
$\textrm{SO}(p)$ for the Riemannian case. Lower-case Latin indices
indicate tensors with respect to diffeomorphisms of
$\mc{M}$. Upper-case Latin indices transform under $G$. Indices for
differential forms will be suppressed when this does not cause
confusion.

We will usually take $\omega$ to be the unique torsion-free spin
connection associated with the frame field $e$, and in this case,
$\omega$ is related to the frame by the Cartan structure equation
\begin{equation}
\label{cart}
\d e^I + \omega^I_J \wedge e^J=0
\end{equation}
However, it is possible to let $e$ and $\omega$ remain independent. In
this case, $\omega$ needs to be determined by other means (see section
(\ref{theories})). The curvature 2-form of $\omega$, $R^I_J[\omega]$
is defined as
\begin{align}
R^I_J[\omega] = \d \omega^I_J + \omega^I_K \wedge \omega^K_J
\end{align}

Averaged quantities will generally be denoted with an overline
(i.e. $\overline{\omega}$) whereas quantities defined by a
compatibility condition with an averaged field will be denoted with
tildes (i.e., $\tilde{e}$).

\section{Details of Averaging Procedure}
In order to carry out the parallel transport procedure alluded to in
the introduction, we will need to choose a connection and a curve. One
possibility would be to choose the metric-compatible Levi-Cevita
connection and parallel transport along the geodesic connecting the
two points (if one exists). A difficulty in this choice is that one is required to
explicitly calculate the equations for the geodesic. Therefore, we
will make another choice, one that does not depend on the choice of
the curve which is naturally available in theories formulated in terms
of a frame field: the {\it Weitzenb\"ock connection}
$W$\cite{weitz}. This is a connection with non-zero torsion, zero
curvature, and zero non-metricity defined via the frame field by
\begin{displaymath}
  W^b_{ca} = e^b_I \partial_a e^I_c
\end{displaymath}

The corresponding parallel transport equation for a vector field $v$ is then
\begin{displaymath}
t^a \partial_a v^b + W^b_{ca} t^a v^c=0
\end{displaymath}
where ${\bf{t}}$ is the tangent vector to the curve along which the parallel transport takes place.  It can be checked by a direct substitution that
\begin{equation}
\label{parasol}
  v^a(x)=P^a_{a'}(x,x') v^{a'}(x')
\end{equation}
solves this equation, where $P^a_{a'}(x,x')=e^a_I (x)
e^I_{a'}(x')$. The object $P^a_{a'}(x,x')$ is the parallel transport
operator that maps a vector at point $x'$ into the parallel
transported vector at reference point $x$. In particular, it can be
noted that $P^a_{a'}(x,x')$ depends only on the initial and final
point and not on the choice of curve connecting the two points; this
reflects the flatness of the Weitzenb\"ock connection $W$ (see
\cite{vdh2010} for additional details).

We now define the average of a tensor field $T^{a_1 \cdots a_n}_{b_1
  \cdots b_m}$ as
\begin{equation}
\label{avgdef}
  \overline{T}^{a_1 \cdots a_n}_{b_1 \cdots b_m} = \frac{1}{V_\Sigma}
\int\limits_\Sigma
P^{a_1}_{a'_1} \cdots P^{a_n}_{a'_n}  P^{b'_1}_{b_1} \cdots P^{b'_n}_{b_n}
 T^{a'_1 \cdots a'_n}_{b'_1 \cdots b'_m}
e' \, \d^p x'
\end{equation}
Here, $\Sigma$ is the region over which the averaging is performed,
$V_\Sigma$ is the volume of that region (as measured by the unaveraged
frame field) and primes refer to dependence on the point $x'$ over
which the integration is performed, and $e'$ is the determinant of the
frame field.

\section{Averaged Manifold}

It can be readily seen from equation (\ref{parasol}) that parallel
transporting the frame field from point $x'$ to point $x$ will simply
give us the frame field at the point $x$.  The corresponding statement
is also true for the metric, that is, the average of the metric is the
metric itself, which is at odds with the intuitive view of averaging
as a smoothing process. Therefore, an alternative process is
necessary. A suggested course of action is to proceed in the following
manner.

One starts the process by averaging the spin connection $\omega$:
\begin{equation}
\label{connavg}
  \overline{\omega}^I_{Ja} = \frac{1}{V_\Sigma}
\int\limits_{\Sigma} e_a^K e^{a'}_K
\omega^I_{Ja'}  e' \d^p x'
\end{equation}
Since $\omega$ is a Lie algebra-valued one-form and since the
averaging procedure ensures that the same is true of
$\overline{\omega}$, one concludes that $\overline{\omega}$ is truly a
connection form.

Given $\overline{\omega}$, can one find a compatible frame field
$\tilde{e}$ with $\overline{\omega}$ as its spin connection? The
Cartan structure equation, equation (\ref{cart}), has an integrability
condition $R^I_J[\overline{\omega}] \wedge \tilde{e}^J=0$ for
$\overline{\omega}$ and $\tilde e$. Under what circumstances is this
condition satisfied? It is clear that a necessary condition is that
$\overline{\omega}$ take values in a subalgebra of the
pseudo-orthogonal Lie algebra given by the metric signature. In
\cite{Schmidt:1973}, it is shown that this condition is sufficient as
well, so the connection determines the metric up to a constant, and
therefore the frame up to a constant and a local Lorentz
transformation.

There are now two possible identifications for the curvature
associated with the averaged manifold.  The first is to calculate the
corresponding curvature $R[\overline{\omega}]$ associated with the
averaged spin connection. The second is to calculate the curvature
$R[\omega]$ and then average to obtain $\overline{R[\omega]}$, which
is generally distinct from $R[\overline{\omega}]$.

If we proceed in the latter manner by assuming that
$\overline{R[\omega]}$ is indeed the curvature of some spin connection
$\tilde \omega$, can one find the corresponding spin connection
$\tilde{\omega}$? The integrability condition is $\overline{R}^J_I
\wedge \tilde{\omega}^K_J - \tilde{\omega}^I_J \wedge \overline{R}^J_K
= \d \overline{R}^K_I$.  This integrability condition does not hold
for {\it all} manifolds, but it is generically
satisfied\cite{Hall:1989}.

What if any relation exists between $\overline{R[\omega]}$ and
$R[\overline{\omega}]$? A partial result is possible once we discuss
how to obtain neighboring averaging regions.  One procedure which
yields neighboring averaging regions of similar size and shape to the
original averaging region is a process coined ``averaging region
coordination'' first described in \cite{Zal} and further illustrated
in \cite{vdh2010,MarsZalaletdinov1997}.  In this procedure, the given
direction vector at $x$, $\xi^d(x)$ is mapped (or coordinated) to a
direction vector for all other points $x'\in\Sigma_x$ through parallel
transport.  The vector
$$P^{d'}_{d}(x,x')\xi^{d}(x)$$
is the image of the given directional vector $\xi^d$ at $x$ to each
point $x'\in \Sigma_x$.  Given an averaging region $\Sigma_x$ at
supporting point $x$ and a direction vector $\xi^d(x)$, one can now
obtain a nearby averaging region $\Sigma_{x+\Delta\lambda\vec\xi}$, by
Lie dragging $\Sigma_x$ a parametric distance $\Delta\lambda$ along
the integral curves of the parallel transported vector field
$P^{d'}_{d}(x,x')\xi^{d}(x)$ for each point $x'\in\Sigma$.  Assuming
this ``averaging region coordination`` one is able to explicitly
calculate the difference in the averaged curvature and the curvature
of the averaged spin connection. See
\cite{Zal,vdh2010,MarsZalaletdinov1997} for details on the
calculation.
\begin{eqnarray*}
R^I_J{}_{cd}[\overline \omega]
&=&-2\overline\omega^I_J{}_{[c,d]} +2\overline\omega^I_K{}_{[c}
\overline\omega^K_{J}{}_{d]}\\
&=&-2\left[ \overline{\omega^I_J{}_{[c,d]}}+\frac{1}{V_\Sigma}\int_{\Sigma}
       {\tt G}^{a}_{[cd]}\omega^I_J{}_{a'}P^{a'}_{a}e'd^px'
       +\overline{\left(\omega^I_J{}_{[c}e_{,d]}\frac{1}{e}\right)}
       -\overline\omega^I_J{}_{[c}\overline{\left(e_{,d]}\frac{1}{e}\right)}\right]
       \\ & \ &\qquad\qquad\qquad+2\overline\omega^I_K{}_{[c}
\overline\omega^K_{J}{}_{d]}\\
&=&\overline{R^I_J{}_{cd}[\omega]}-2\overline{\omega^I_K{}_{[c}
\omega^K_J{}_{d]}} +2\overline\omega^I_K{}_{[c}\overline\omega^K_{J}{}_{d]}
       \\ & &\qquad\qquad-\frac{2}{V_\Sigma}\int_{\Sigma}
       {\tt G}^{a}_{[cd]}\omega^I_J{}_{a'}P^{a'}_{a}e'd^px'
       -2\overline{\left(\omega^I_J{}_{[c}e_{,d]}\frac{1}{e}\right)}
       +2\overline\omega^I_J{}_{[c}\overline{\left(e_{,d]}\frac{1}{e}\right)}
 \end{eqnarray*}
where the term containing ${\tt G}^a_{bd}$ is the result of changes
in the parallel transporter as one changes position.  Explicitly one writes
$${\tt G}^a_{bd}=P^a_{a'}(P^{a'}_{b,d}+P^{a'}_{b,d'}P^{d'}_{d}).$$
Using the definitions for both $P^a_{a'}$ and $W^a_{bd}$ one observes
that 
\begin{displaymath}
{\tt
  G}^a_{bd}=W^a_{bd}-P^a_{a'}P^{b'}_{b}P^{d'}_{d}W^{a'}_{b'd'}
\end{displaymath}
which is just the difference between the Weitzenb\"ock connection at $x$ and
the image of the Weitzenb\"ock connection at $x'$ parallel transported
to $x$.  The two terms containing the $e_{,d}$ are the result of
changes in the volume as one changes position.

\subsection{Change of averages under change of base point}

While a full investigation is beyond the scope of this paper, we will
here show that when the base point $x$ of $\Sigma_x$ is shifted, on
average the average of a scalar quantity $f$ changes more slowly than
the corresponding unaveraged quantity.

Let us start with the definition of averaging for a scalar function $f$ on
$\mc{M}$:

\begin{displaymath}
  \bar{f} = \frac{1}{V}\int\limits_{\Sigma_x} f \d V 
\end{displaymath}

\ni where $\d V$ is the volume element associated with the frame field $e$
and $V=\int_{\Sigma_x} \d V$ is the volume of $\Sigma_x$. Let us now shift the region of
integration $\Sigma_x$ by an amount $\delta x$ to a new region
$\Sigma_{x'}$ with a base point at $x'$.  We assume $\delta x$ to be
infinitesimal. In addition, we will need to assume that the map $x'^a
= x^a+\delta x^a$ is volume-preserving.

\begin{figure}[h]
  \centering
\includegraphics[height=55mm]{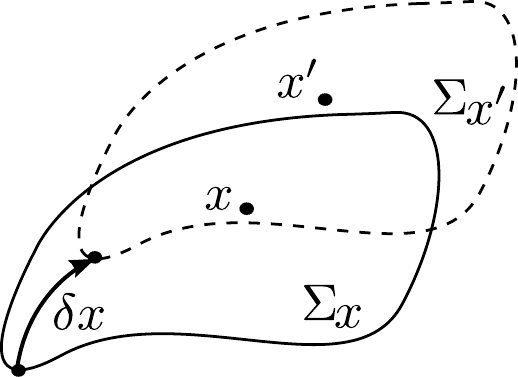}
\caption{$\Sigma_x$ being transformed into $\Sigma_{x'}$ by the map
$x'^a = x^a+\delta x^a$. }
\end{figure}

A calculation leads to the induced shift $\delta \bar{f}$ of the average 
value $\bar{f}$ being given by

\begin{displaymath}
 \delta \bar{f} = \frac{1}{V} \int\limits_{ \Sigma_x}
\partial_a (f(x) \delta x^a) n_a \d S
\end{displaymath}

\ni By comparison, the corresponding change $\delta f$ in the
unaveraged quantity $f$ is given by

\begin{displaymath}
  \delta f = \delta x^a \partial_a f
\end{displaymath}

It then follows from the triangle inequality and the assumptions stated
above that

\begin{displaymath}
\vert \delta \bar{f} \vert  \leq \left| \overline{\delta f} \right|
\end{displaymath}

\ni where the average on the right-hand side is taken over $\Sigma_x$.
This equation can be interpreted as stating that on average, the 
averaged $f$ changes more slowly than the unaveraged $f$.

\section{An example}

Let us as an example explicitly perform the averaging for the manifold
in volume preserving coordinates to eliminate the effects of changes
in the volume.  Consider the manifold $(\mathbb{R}^2,\d s^2)$, with
$\d s^2 = f(x)^{-2}\d x^2 + f(x)^2 \d y^2$. The line element yields
the ortho-normal frame field $e^1 = f(x)^{-1}\d x$, $e^2 = f(x) \d
y$. The corresponding nontrivial components of the Weitzenb\"ock connection are
\begin{displaymath}
W^{x}_{xx}=-\frac{f'(x)}{f(x)}\qquad W^{y}_{yx}=\frac{f'(x)}{f(x)}
\end{displaymath}
From the Cartan structure equation (\ref{cart}), we obtain
\begin{displaymath}
  \omega^1_2 = - f(x)f'(x) \d y =- \frac{1}{2}([f(x)]^2)' \d y
\end{displaymath}
as the only non-zero component. If we pick $\Sigma$ to be a square with side coordinate length $2\delta$ centred at $(x,y)$, equation
(\ref{connavg}) now tells us that
\begin{align}
    \overline{\omega}^1_{2}
    &=\left[\frac{1}{4 \delta^2 }\int\limits_{\Sigma} e_y^K e^{a'}_K
\omega^1_{2a'}  e' \d^2 x'\right] \d y\nonumber \\
    &=\left[\frac{1}{4 \delta^2}\int\limits_{y-\delta}^{y+\delta}
\int\limits_{x-\delta}^{x+\delta} \frac{f(x)}{f(x')} \left( - f'(x') f(x')\right)
\d x' \d y'\right] \d y\nonumber \\
    &=\left[-\left(\frac{f(x+\delta)-f(x-\delta)}{2\delta}\right)f(x)\right] \d y
\end{align}
It is a simple exercise in taking limits to show that as $\delta
\rightarrow 0$, $\overline{\omega}^1_{2} \rightarrow - f'(x) f(x) \d y$ as
it should, since in that limit $\overline{\omega}$ should reduce to
the unaveraged connection $\omega$.

The next step is to find a frame $\tilde{e}$ compatible with
$\overline{\omega}$.  Again, we use equation (\ref{cart}) and this
time we find the system of equations
\begin{align}
\label{cartsys}
  \begin{cases}
    \d \tilde{e}^1  + \overline{\omega}^1_2 \wedge \tilde{e}^2 = 0 \\
    \d \tilde{e}^2  + \overline{\omega}^2_1 \wedge \tilde{e}^1 = 0
  \end{cases}
\end{align}
Assume that $\tilde{e}^1 = F(x) \d x$ for some as yet undetermined
function $F(x)$. This leads to
\begin{align}
  - \frac{f(x)}{2 \delta} \left( f(x+\delta) - f(x-\delta) \right) \d
  y \wedge \tilde{e}^2 = 0
\end{align}
which implies that $\tilde{e}^2=h(x,y) \d y$. Inserting this into the
second equation in (\ref{cartsys}) gives
\begin{align}
  \frac{\partial h}{\partial x} \d x \wedge \d y +
\frac{ f(x) }{2 \delta} \left( f(x+\delta) - f(x-\delta) \right)F(x)
 \d x \wedge \d y = 0
\end{align}
which in turn shows that
\begin{align}
  h(x,y) = h_0(y) + \frac{1}{2 \delta} \int f(x)\left( f(x+\delta) -
    f(x-\delta) \right)F(x) \d x
\end{align}
for some function $h_0(y)$. By requiring that $h(x,y) \rightarrow
f(x)$ as $\delta \rightarrow 0$, a small calculation shows that
$h_0(y)=0$ and $F(x)=f(x)^{-1}$ resulting in a frame field of the form
\begin{align}
  \tilde{e}^1 = f(x)^{-1} \d x \qquad\qquad \tilde{e}^2 =
  \left[\frac{1}{2\delta}\int (f(x+\delta)-f(x-\delta))dx\right] \d y
\end{align}

As for the curvatures, we start by calculating $R^I_J[\omega]$
\begin{equation}
  R^I_J[\omega]= - \left[ f(x) f'(x) \right]' \d x \wedge \d y =
- \frac{1}{2}([ f(x) ]^2)'' \d x \wedge \d y
\label{R}
\end{equation}
with all the other components of $R^I_J[\omega]$ equal to
zero. Averaging, we find
\begin{equation}
    \overline{R}^1_{2}(\omega) =-\left[
\frac{f(x+\delta)f'(x+\delta)-f(x-\delta)f'(x-\delta)}{2\delta}\right]\d x \wedge \d y
\label{R_BAR}
\end{equation}
We can also compute the curvature of the averaged connection
$\overline{\omega}$; this leads to
\begin{equation}
    {R}^1_{2}(\overline\omega) =
-\left[\frac{f(x)f'(x+\delta)-f(x)f'(x-\delta)+f(x+\delta)f'(x)-
f(x-\delta)f'(x)}{2\delta}\right]\d x \wedge \d y
\end{equation}
One can show that the two curvatures both approach $- \left[ f(x)
  f'(x) \right]' \d x \wedge \d y =R^1_2[\omega]$ in the limit $\delta
\rightarrow 0$.

Two-dimensional spaces of constant curvature are prescribed 
if $f(x)=\sqrt{1-kx^2}$ for $k=-1,0,1$.
Substitution of this expression into equations (\ref{R}) and 
(\ref{R_BAR}) shows that
$\overline{R}{}^1_{2}(\omega)={R}^1_{2}(\omega)=k \ \d x \wedge \d y$, 
as might be expected for spaces of constant curvature.  In contrast, 
${R}^1_{2}(\overline\omega)\not={R}^1_{2}(\omega)$, which 
suggests that the appropriate primary object to consider when averaging 
or smoothing manifolds is the curvature and not the
connection.

\section{Averaging and gravitational theories}
\label{theories}

In physical applications, $e$ and $\omega$ will ultimately be given by
the field equations of some geometric gravitational theory, along with
some boundary conditions. If $\omega$ is assumed to be the spin
connection compatible with $e$, we may for instance choose the GR
action in Palatini form, since the compatibility of $e$ and $\omega$
then falls out naturally as a field equation. On the other hand, if
$e$ and $\omega$ are a priori assumed to be independent, it is natural
to consider other theories of gravity, such as Einstein-Cartan theory
and its Poincar\'e Gauge Theory cousins \cite{Hehl:1994ue}. For an
 overview of actions for geometric gravitational
theories, see \cite{Peldan:1993hi}.  For instance, for general
relativity the action would take the form
\begin{displaymath}
  S=\int e^a_I e^b_J R_{ab}^{IJ}[\omega] e \, \d^p x
\end{displaymath}
Varying this action with respect to the frame $e_{aI}$ gives the usual
vacuum Einstein equation $R_{ab}=0$. If we consider $e$ and $\omega$
to be independent, varying with respect to the connection $\omega$
will give the compatibility condition (\ref{cart}).

No matter which gravitational theory is considered, the field
equations for the unaveraged $e$ and $\omega$ will determine equations
and integrability conditions for the corresponding averaged
quantities. The form of these equations will be studied in a future
paper.

\end{document}